\documentclass[12pt]{IEEEtran}
\usepackage{amsmath,amssymb,verbatim}
\usepackage{graphicx}

\newtheorem{theorem}{Theorem}
\newtheorem{lemma}{Lemma}
\newtheorem{proposition}{Proposition}
\newtheorem{definition}{Definition}
\newtheorem{remark}{Remark}
\newtheorem{example}{Example}

\newcommand{\enproof}{\hfill $\Box$ \vspace*{1ex}}
\newcommand{\enlem}{\hfill $\Diamond$ \end{lemma}}
\newcommand{\closedef}{\hfill $\Diamond$ \end{definition}}
\newcommand{\enth}{\hfill $\Diamond$ \end{theorem}}
\newcommand{\encor}{\hfill $\Diamond$ \end{corollary}}
\newcommand{\enprop}{\hfill $\Diamond$ \end{proposition}}
\newcommand{\encond}{\hfill $\Diamond$ \end{condition}}
\newcommand{\myQED}{\hfill $\Box$}

\newcommand{\exam}[1]{\begin{example}\label{ex:#1}}
\newcommand{\enexam}{\enproof}
\newcommand{\beremark}[1]{\begin{remark}\label{rmk:#1}}
\newcommand{\enremark}{\enproof}

\newcommand{\mymathbb}[1]{{\mathbb #1}} 
\newcommand{\mymathsf}[1]{{\mathsf #1}} 

\newcommand{\cA}{{\cal A}}
\newcommand{\sA}{\mymathsf{A}}
\newcommand{\cB}{{\cal B}}
\newcommand{\sB}{\mymathsf{B}}

\newcommand{\cC}{{\cal C}}
\newcommand{\bC}{\mymathbb{C}}
\newcommand{\cD}{{\cal D}}
\newcommand{\cE}{{\cal E}}

\newcommand{\sF}{\mymathsf{F}}
\newcommand{\myF}{{\sF}}

\newcommand{\sH}{\mymathsf{H}}
\newcommand{\cI}{{\cal I}}

\newcommand{\sL}{\mymathsf{L}}
\newcommand{\cM}{{\cal M}}

\newcommand{\sN}{\mymathsf{N}}

\newcommand{\sP}{\mymathsf{P}}
\newcommand{\cQ}{{\cal Q}}

\newcommand{\cR}{{\cal R}}

\newcommand{\cT}{{\cal T}}
\newcommand{\cU}{{\cal U}}

\newcommand{\cX}{{\cal X}}

\newcommand{\bZ}{\mymathbb{Z}}

\renewcommand{\phi}{\varphi}
\renewcommand{\subset}{\subseteq}
\renewcommand{\tilde}{\widetilde}
\renewcommand{\hat}{\widehat}

\newcommand{\mbm}[1]{\mbox{\boldmath $#1$}}
\newcommand{\mateq}{\stackrel{\rm m}{\sim}}

\newcommand{\cmple}{^{\rm c}}

\newcommand{\field}{\mymathbb{F}}

\newcommand{\Erc}{E_{\rm r}}
\newcommand{\Wch}{W}

\newcommand{\tnsr}{\otimes}

\newcommand{\lag}{\langle}
\newcommand{\rag}{\rangle}
\newcommand{\crd}[1]{|#1|}
\newcommand{\bra}[1]{\lag #1 |}
\newcommand{\ket}[1]{| #1 \rag}
\newcommand{\indc}{{\bf 1}}
\newcommand{\syp}[2]{( #1,  #2 )_{\rm sp}}
\newcommand{\spn}{\mymathsf{span}}

\newcommand{\dmn}{d}
\newcommand{\Hch}{{\sH}}
\newcommand{\Hgn}{{\sH}}
\newcommand{\Id}{{\rm I}}
\newcommand{\Bop}{\sL}
\newcommand{\Hcd}{\cC}
\newcommand{\Bcd}{\cB}

\newcommand{\Fav}{F_{\rm a}}
\newcommand{\Pbad}{G}
\newcommand{\kpr}{m} 
\newcommand{\imu}{{\rm i}}
\newcommand{\Ebasis}{\sN}
\newcommand{\Ebe}{N}
\newcommand{\Ecl}{E_{\rm cl}}
\newcommand{\Icr}{J}
\newcommand{\Fbar}{\overline{F}}
\newcommand{\Aso}{\sA}  
\newcommand{\Acn}{\sA}  
\newcommand{\Bcn}{\sB}  
\newcommand{\CPex}{\cM}
\newcommand{\CPexO}{M}
\newcommand{\Cso}{L}

\newcommand{\ahat}{\hat{a}}
\newcommand{\varvn}{u}
\newcommand{\varx}{s}

\newcommand{\tnsn}{^{\tnsr n}}

\title{
Lower Bounds on the Quantum Capacity\\ 
and Highest Error Exponent of\\
General Memoryless Channels}

\author{Mitsuru Hamada\\[1ex]
{\normalsize Quantum Computation and Information Project (ERATO)\\
     Japan Science and Technology Corporation\\
              201 Daini Hongo White Bldg.,\
      5-28-3, Hongo, Bunkyo-ku, Tokyo 113-0033, Japan\\
E-mail: {\tt mitsuru@ieee.org}}
}
\date{Feb.\ 1, 2002}

\begin{document} 


\maketitle 

\begin{abstract}
Tradeoffs between the information rate and 
fidelity of quantum error-correcting codes are discussed.
Quantum channels to be considered are
those subject to independent errors
and modeled as tensor products of copies of a 
general completely positive linear map,
where the dimension of the underlying Hilbert space is a prime number.
On such a quantum channel, the highest fidelity
of a quantum error-correcting code of length $n$ and rate $R$
is proven to be lower bounded by
$1 - \exp [-n E(R) + o(n)]$ for some function $E(R)$.
The $E(R)$ is positive below some threshold $R_0$,
a direct consequence of which is 
that $R_0$ is a lower bound on the quantum capacity.
This is an extension of the author's previous result
[M.~Hamada, Phys.\ Rev.\ A, vol.\ 65, 052305, 2002;
LANL e-Print, quant-ph/0109114, 2001].
While it states the result for the depolarizing channel
and a slight generalization of it (Pauli channels),
the result of this work applies to general discrete memoryless channels,
including channel models derived from a physical law of time evolution.
\end{abstract}



\section{Introduction \label{ss:intro}}

Quantum error-correcting codes 
(also called quantum codes or {\em codes}\/ in this work)
have attracted much attention
as schemes that protect
quantum states from decoherence during quantum computation.
Shor invented the first code and
stated that the ultimate goal would be to define the quantum
analog of Shannon's channel capacity, and find encoding schemes
which approach this capacity~\cite{shor95}.
On quantum memoryless channels, several bounds on the quantum capacity
are known~\cite{shor95,schumacher96,bennett96m,dss98,barnum00}.
Good surveys on this problem are given in the introductory section of
\cite{barnum00} and in \cite{holevo01s}.
There is a conjecture that the known upper bound based on the notion called
coherent information is tight~\cite{barnum98}, \cite[Section~VI]{barnum00}.
On the other hand, the existing lower bounds seem to have left much room
for improvement. For example, there is a lower bound 
on the capacity of the so-called depolarizing channel
which can be proved by a random coding argument that evaluates
the average performance over the whole ensemble of standard 
quantum error-correcting codes~\cite{gottesmanPhD,preskillLN},
or by an argument using 
an entanglement purification protocol~\cite{bennett96m}.
Shor and Smolin~\cite{ss97,dss98} argued that this bound
is not tight showing the existence of quantum codes,
which are, in a sense, analogous to
classical concatenated codes~\cite{forney},
of performance beyond it for a limited class of very noisy channels. 
The present author recently strengthened the result on
standard quantum error-correcting codes~\cite{gottesmanPhD,preskillLN}
in another direction, namely, established
exponential convergence of fidelity of codes
used on slight generalizations of the depolarizing channel~\cite{hamada01e}.
In other words, using these simple channels, he illustrated that certain
results and ideas around the error exponent problem in classical information theory, 
which has been a central issue%
~\cite{gallager65,gallager,csiszar_koerner,litsyn99,ash,viterbi_omura,slepianIT}, can be extended to quantum channels.
The classical error exponent problem is, 
roughly speaking, to determine the function
$\Ecl(R,W)$
such that the decoding error probability ${\rm P}^{\star}_{n}$ of 
the best code of length $n$ and rate $R$ behaves like
${\rm P}^{\star}_{n} \approx \exp [ -n \Ecl(R,W) ]$ on a channel $W$.
The $\Ecl(R,W)$, which is called the reliability function
or the highest achievable error exponent of a channel $W$,
is positive below the capacity of $W$, and decreasing in $R$.
See, e.g., \cite{gallager,csiszar_koerner}
for precise definitions of the reliability function, 
\cite{litsyn99} for a recent development, 
and \cite{slepianIT,gallager01} for history.
There is no reason to employ codes of rates near the capacity exclusively
because the less $R$ is, the greater $\Ecl(R,W)$ is, and hence
the less ${\rm P}^{\star}_{n} \approx \exp [ -n \Ecl(R,W) ]$ is exponentially.

The goal of this work is to show such exponential convergence
of the fidelity of quantum error-correcting codes
on a much wider class of channels.
The channels to be considered here are those subject to independent errors
and modeled as tensor products of copies of a 
general completely positive (CP) linear map~\cite{kraus71,choi75}.
Our channel class includes those derived from a physical law of time evolution,
or from master (Lindblad) equations~\cite{preskillLN,AlickiLendi,nielsen_chuang,GardinerZoller}, though it is stipulated 
that the Hilbert spaces underlying channels 
have dimensions of prime numbers.
One example of such channels
is the amplitude-damping channel,
which has often been discussed in the context of quantum error correction~\cite{preskillLN,nielsen_chuang,KnillLaflamme97}.
Despite the fact that this channel has often been treated as a model of quantum
noise suffered during quantum computation, it has been not known
whether standard quantum error-correcting codes work reliably
at a positive rate for all large enough code lengths
on this channel. 
%

This work was inspired by
Matsumoto and Uyematsu~\cite{MatsumotoUyematsu01},
who tried to prove a lower bound
on the quantum capacity of a general memoryless channel
using standard quantum error-correcting codes.
However, their proof turned out to be wrong unfortunately
[R.~Matsumoto and
T.~Uyematsu, 24th Symposium on Information Theory and Its Applications,
Kobe, Hyogo, Japan, Dec.~7, 2001].
In fact, they used the inequality
similar to that in Lemma~5 below, which allegedly 
held for the standard fidelity measure
(minimum fidelity, denoted by $F(\Hcd)$ in this paper), 
in \cite{MatsumotoUyematsu01},
but this fails as shown in Example 3 below.
Moreover, their bound~\cite{MatsumotoUyematsu01}
is smaller than Preskill's lower bound~\cite[Section~7.16.2]{preskillLN}
for the so-called Pauli channels in general.
It may be said that their contribution lies in the use of the estimate due
to Calderbank {\em et al.},\/ 
which will be given in Lemma~\ref{lem:C_uniform} below in a slightly different form,
in the present context. This is what this work has inherited
from \cite{MatsumotoUyematsu01}.
Thus, the question of whether quantum error-correcting codes
work reliably on general channels or not is yet to be answered,
which this paper is concerned with
from an information-theoretic viewpoint.
Specifically, exponential convergence of
the fidelity of codes on general memoryless channels is established.
The proof to be presented below exploits existing information-theoretic
techniques, such as the method of types%
~\cite{csiszar_koerner,csiszar98,CsiszarKoerner81a,cover_th},
as well as a previously unused property of standard
quantum-error-correcting codes.

We remark that in the setting where
classical messages are sent over quantum channels,
the error exponent problem has been discussed by Burnashev and Holevo%
~\cite{BurnashevHolevo97} and Holevo~\cite{holevo00}
while this paper is concerned with
the problem of preserving or transmitting quantum states
in the presence of quantum noise. 
Note also that 
the error exponents of quantum error-{\em detecting}\/ codes,
which do not correct errors but only detect errors,
have been discussed by Ashikhmin {\em et al.}~\cite{ABargKnillL00}.

The rest of the paper is organized as follows.
Section~II presents the main result.
In Section~III, a performance measure for codes, 
which is called the minimum average fidelity, 
is introduced and it is argued that
evaluating this measure gives a good estimate for the standard fidelity.
Section~IV reviews the standard quantum codes,
and Section~V gives bounds on the minimum average fidelity of codes. 
Finally, the main result is proved in Section~VI,
which is followed by a concluding section.
Appendices are given to prove a proposition, two lemmas,
and an inequality between the proposed bound and
the previously known one.

\section{Main Result \label{ss:main_result} }

As usual,
all possible quantum operations
and state changes,
including 
quantum channels, 
are described in terms of 
{\em completely positive}\/ (CP) linear maps%
~\cite{kraus71,choi75,schumacher96,barnum00}.
In this work, only 
{\em trace-preserving completely positive}\/ (TPCP) linear maps are treated.
Given a Hilbert space $\Hgn$ of finite dimension,
let $\Bop(\Hgn)$ denote the set of linear operators on $\Hgn$. 
In general, every CP linear map $\CPex: \Bop(\Hgn) \to \Bop(\Hgn)$ 
has an operator-sum representation
$\CPex(\rho) = \sum_{i\in\cI} \CPexO_i \rho \CPexO_i^{\dagger}$ with some
set of operators $\{ \CPexO_i \in\Bop(\Hgn) \}_{i\in\cI}$, 
which is not unique~\cite{choi75,schumacher96}.
When $\CPex$ is specified by a set of operators
$\{ \CPexO_i \}_{i\in\cI}$ 
in this way, we write
$\CPex \sim \{ \CPexO_i \}_{i\in\cI}$. 
Note that we can always have $\crd{\cI}$ equal to $(\dim \Hch)^2$,
including null operators in $\{ \CPexO_i \}_{i\in\cI}$
if necessary~\cite{choi75}.

Hereafter, $\Hch$ denotes an arbitrarily fixed Hilbert space
of dimension $\dmn$, which is a prime number.
A quantum channel is a sequence of
TPCP linear maps 
$\{ \cA_n  : \Bop(\Hch^{\tnsr n}) \to \Bop(\Hch^{\tnsr n}) \}$;
the map $\cA_n$ with a fixed $n$ is also called a channel.
We want a large subspace $\Hcd \subset \Hch^{\tnsr n}$
every state vector in which remains almost unchanged 
after the effect of a channel
followed by the action of some suitable recovery process.
The recovery process is again described as a TPCP linear map
$ 
\cR: \Bop(\Hch^{\tnsr n}) \to \Bop(\Hch^{\tnsr n}). 
$ 
A pair $(\Hcd, \cR)$ consisting of such a subspace $\Hcd$
and a TPCP linear map $\cR$
is called a {\em code}\/
and its performance is evaluated in terms of the minimum fidelity%
~\cite{KnillLaflamme97,dss98,barnum00}
\begin{equation}\label{eq:minimum_fidelity}
F(\Hcd, \cR\cA_n) = \min_{ \ket{\psi} \in \Hcd } 
\lag\psi| \cR\cA_n(|\psi\rag \lag\psi|) |\psi\rag,
\end{equation}
where $\cR\cA_n$ denotes the composition of $\cA_n$ and $\cR$.
Throughout, bras $\bra{\cdot}$ and kets $\ket{\cdot}$ are
assumed normalized.
Sometimes, a subspace $\Hcd$ alone is called a code
assuming implicitly some recovery operator.
Let $F_{n,k}^{\star}(\cA_n)$ denote the
supremum of
$F(\Hcd, \cR\cA_n)$ such that there exists a code $(\Hcd, \cR)$
with $\log_{\dmn} \dim \Hcd \ge k$, where $n$ is a positive integer
and $k$ is a nonnegative real number.
This paper gives an
exponential lower bound on $F_{n,k}^{\star}(\cA_n)$,
where for simplicity
we state the result in the case where the channel is memoryless, 
i.e., when $\cA_n = \cA^{\tnsr n}$ for some 
$\cA : \Bop(\Hch) \to \Bop(\Hch)$;
the channel $\{ \cA_n = \cA^{\tnsr n} \}$ is referred to as
the memoryless channel $\cA$.

The codes to be proven
to have the desired performance are
{\em symplectic (stabilizer or additive) codes}\/%
~\cite{gottesman96,crss97,crss98,knill96a,knill96b,rains99}.
In designing these codes, the following basis of $\Bop(\Hch)$,
which has some nice algebraic properties, is used.
Fix an orthonormal basis (ONB) 
$\{ |0\rag,\dots, |\dmn-1\rag \}$ of $\Hch$. 
The `error basis' is 
$\Ebasis=\{ \Ebe_{(i,j)}=X^i Z^j \}_{(i,j)\in\cX}$
where $\cX=\{0,\dots, \dmn-1\}^2$ and
the unitary operators $X, Z \in \Bop(\Hch)$ are defined by
\begin{equation}\label{eq:error_basis}
X |j \rag  = |(j-1) \bmod \dmn \, \rag, \quad
Z |j \rag = \omega^ j |j \rag
\end{equation}
with $\omega$ being a primitive $\dmn$-th root of unity~\cite[Section~IV-15]{weyl31}.
When $\dmn=2$, the basis elements become $I, X, XZ, Z$,
which are the same as the identity and three Pauli operators
up to a phase factor.
As usual,
the classical Kullback-Leibler information
(informational divergence or relative entropy)
is denoted by $D$ and entropy by $H$~\cite{csiszar_koerner,csiszar98,cover_th}.
Specifically, for probability distributions $P$ and $Q$ on a finite set $\cX$,
we define $D(P||Q)$ by $D(P||Q)=\sum_{x\in\cX} P(x) \log_{\dmn} [P(x)/Q(x)]$ and
$H(Q)$ by $H(Q)= - \sum_{x\in\cX} Q(x) \log_{\dmn} Q(x)$.
By convention, we assume 
$\log(a/0) = \infty$ for $a>0$ and $0\log 0= 0 \log (0/0) = 0$. 

To state our result, we associate a probability distribution with  a channel.
\begin{definition}\label{def:P4A}
For a memoryless channel $\cA: \Bop(\Hch) \to \Bop(\Hch)$,
we define a probability distribution $P_{\cA}=P_{\cA,\Ebasis}$ on $\cX$ as follows.
For an operator-sum representation $\cA \sim \{ A_u \}_{u\in\cX}$, 
expand $A_u$ in terms of the error basis $\Ebasis$ as
$A_u = \sum_{v\in\cX} a_{uv} \Ebe_v$, $u\in\cX$.
Then,
\[
P_{\cA}(v)=P_{\cA,\Ebasis}(v)= \sum_{u\in\cX} |a_{uv}|^2, \quad v\in\cX.
\]
\closedef

{\em Remarks:}\/
With $\cA$ and $\Ebasis$ fixed,
the $P_{\cA}$ does not depend on the choice of $\{ A_u \}_{u\in\cX}$
while it depends on $\Ebasis$ as well as $\cA$.
That $\sum_{v\in\cX} P(v)=1$ readily follows
from the trace-preserving condition
$\sum_{u\in\cX} A_u^{\dagger} A_u =I$ and the property of the basis $\Ebasis$
that $\Ebe_u^{\dagger} \Ebe_v = I$ if and only if $u=v$~\cite{knill96a}.
\enremark 

This paper's main result is the following one. 
\begin{theorem}\label{th:main}
Let integers $n$, $k$ and
a real number $R$ satisfy $0 \le k \le \lceil Rn \rceil $ and $0 \le R \le 1$ 
(a typical choice is $k = \lceil Rn \rceil$ for an arbitrarily 
fixed rate $R$).
Then, for any memoryless channel $\cA: \Bop(\Hch) \to \Bop(\Hch)$,
and for any choice of the basis 
$\{ |0\rag,\dots, |\dmn-1\rag \}$ and $\omega$ which determine $\Ebasis$, 
we have 
\[
 F_{n,k}^{\star}(\cA^{\tnsr n}) \ge 1 - 2\dmn^2(n+1)^{2(\dmn^2-1)} \dmn ^{ - n E(R,P_{\cA,\Ebasis}) }
\]
where
\[
E(R,P)=\min_{Q} [ D(Q||P) + |1-H(Q)-R|^+ ],
\]
$|x|^+ =\max\{x,0\}$,
the minimization with respect to $Q$ is
over all probability distributions on $\cX=\{0,\dots,\dmn-1\}^2$.
\enth

An immediate consequence of the theorem is that
the quantum capacity~\cite{schumacher96,bennett96m,dss98,barnum00} 
of $\cA$ is lower bounded by 
\begin{equation}\label{eq:lb2cap}
\max_{\Ebasis}[1-H(P_{\cA,\Ebasis})],
\end{equation}
where the maximum is over all choices of the basis
$\{ |0\rag,\dots, |\dmn-1\rag \}$ of $\Hch$ and the primitive $\dmn$-th
root of unity $\omega$.
To be precise,
the capacity of $\{ \cA_n \}$ is defined as 
the supremum of achievable rates on $\{ \cA_n \}$,
where a rate $R$ is said to be achievable if there exists
a sequence of codes $\{ (\Hcd_n, \cR_n) \}$
such that
$\liminf_{n} \log_{\dmn} {\dim} \Hcd_n / n \ge R$
and
$\lim_{n} F(\Hcd_n, \cR_n\cA_n) = 1$.%
\footnote{In the literature, $\liminf_{n} \log_{\dmn} {\dim} \Hcd_n / n \ge R$ 
is sometimes replaced by $\limsup_{n} \log_{\dmn} {\dim} \Hcd_n / n \ge R$
(e.g., \cite{barnum00}). 
Note also that in the definition of the quantum capacity 
(for transmission of subspaces)
by Barnum {\em et al.}~\cite{barnum00}, 
a slightly more general setting is assumed, 
i.e., two Hilbert spaces $H_s$ and $H_c$ are used instead of $\Hch$,
but our bound is also valid in their setting because
we can put $H_s=H_c=\Hch$.
Apart from this difference, there is a seemingly different definition
of the quantum capacity using entanglement fidelity, but actually they
are the same~\cite{barnum00}.}
To see the bound,
observe that $E(R,P)$ is positive for $R<1-H(P)$
due to the basic inequality $D(Q||P) \ge 0$ where equality occurs
if and only if $Q=P$~\cite{csiszar_koerner}.
The bound $1-H(P_{\cA})$ appeared earlier
in Preskill~\cite[Section~7.16.2]{preskillLN}
in the case where $d=2$ and $(a_{uv})$ is diagonal.
The restriction of $(a_{uv})$ being diagonal
also exists in this author's previous result~\cite{hamada01e}.
Namely, it treated channels of the form 
$\cA \sim \{ \sqrt{P(u)} \Ebe_{u} \}_{u\in\cX}$ with some probability
distribution $P$ on $\cX$, which are sometimes called Pauli channels
especially for $\dmn=2$.


Another direct consequence of the theorem is
\begin{equation}\label{eq:new_exp}
\liminf_{n\to\infty} - \frac{1}{n}\log_{\dmn} [1-F^{\star}_{n,Rn}(\cA\tnsn)] 
\ge \max_{\Ebasis} E(R,P_{\cA,\Ebasis}), 
\end{equation}
where the range of the maximization is the same as that for 
(\ref{eq:lb2cap}) above.
This bound resembles the random coding exponent $\Erc(R,\Wch)$
of a classical channel $\Wch$. 
As mentioned in \cite{hamada01e},
the function $E(R,P)$ is, in fact,
the `slided' random coding exponent $\Erc(R+1,\Wch)$ 
of some simple classical channel $\Wch$, i.e.,
the additive channel defined by $\Wch(y|x)=P(y-x)$, $x,y\in\cX=\bZ/\dmn\bZ$,
which becomes the quaternary (completely) symmetric channel~\cite{mceliece}
in the case where $\dmn=2$ and $\cA$ is the depolarizing channel.
In \cite{hamada01e}, one can find
another form of $E$, which is the translation of
an older form of classical random coding exponent $\Erc$
known in the literature
(see, e.g., \cite{csiszar_koerner}, pp.~168, 192--193, 
and \cite{gallager65,gallager})
and suitable for computing $E(R,P_{\cA,\Ebasis})$ numerically
(Fig.~\ref{fig:1}; also Fig.~1 of \cite{hamada01e}).

It should be remarked that, for the obvious reason,
the bounds in (\ref{eq:lb2cap})
and (\ref{eq:new_exp}) actually can be replaced by
\[
\max_{\cU,\Ebasis} [1-H(P_{\cU\cA,\Ebasis})] \quad \mbox{and}\quad
\max_{\cU,\Ebasis} E(R,P_{\cU\cA,\Ebasis}), 
\]
where $\cU\cA$ denotes the composition of $\cA$ and $\cU$,
the map $\cU$ ranges over all TPCP ones on $\Bop(\Hch)$,
and the range of $\Ebasis$ is the same as above.
The role of $\cU$ is preprocessing before the recovery
operation $\cR$, so that restricting the range of $\cU$ to
the set of easily implementable ones, say, to that of
all unitary maps of the form $\cU(\rho)=U \rho U^{\dagger}$
with some unitary operator on $\Hch$, may be reasonable.

In the case of the depolarizing channel,
the relationship between this paper's bound (or that of \cite{hamada01e})
and the previously known bounds are best understood with Fig.~\ref{fig:1},
which depicts $E(R,P_{\cA,\Ebasis})=E(R,P)=E(R,p)$ in the case where $\dmn=2$
and $P\big((0,0)\big)=1-p$, $P(u)=p/3$ for $u \ne (0,0)$, $u \in \cX = \{ 0,1 \}^2$ with $p=1.5\times 10^{-3}j$, $j=0,1,\dots$.
This applies to the depolarizing channel
$\cA \sim \{\sqrt{1-p} \, I,  \sqrt{p/3}\, X,
\sqrt{p/3}\, XZ, \sqrt{p/3}\, Z \}$.
For this channel,
the known bound $1-H_1(p)$~\cite[Fig.~8]{bennett96m}, \cite{gottesmanPhD,preskillLN}, where
\[
H_1(p)=-p\log_2(p)-(1-p)\log_2(1-p)+p\log_{2} 3,
\]
appears in Fig.~\ref{fig:1} as the curve
on which the surface $E(R,p)$ meets the horizontal $pR$-plane.
The Shor-Smolin code~\cite{ss97,dss98}
has improved this lower bound slightly for a limited range of $p$ around the 
point $(p^{\star}, 0, 0)$, where the lower bound $1-H_1(p)$ vanishes
[$1-H_1(p^{\star})=0$, $p^{\star}\approx 0.1893$].
\begin{figure}
\begin{center}
\includegraphics[scale=1.0]{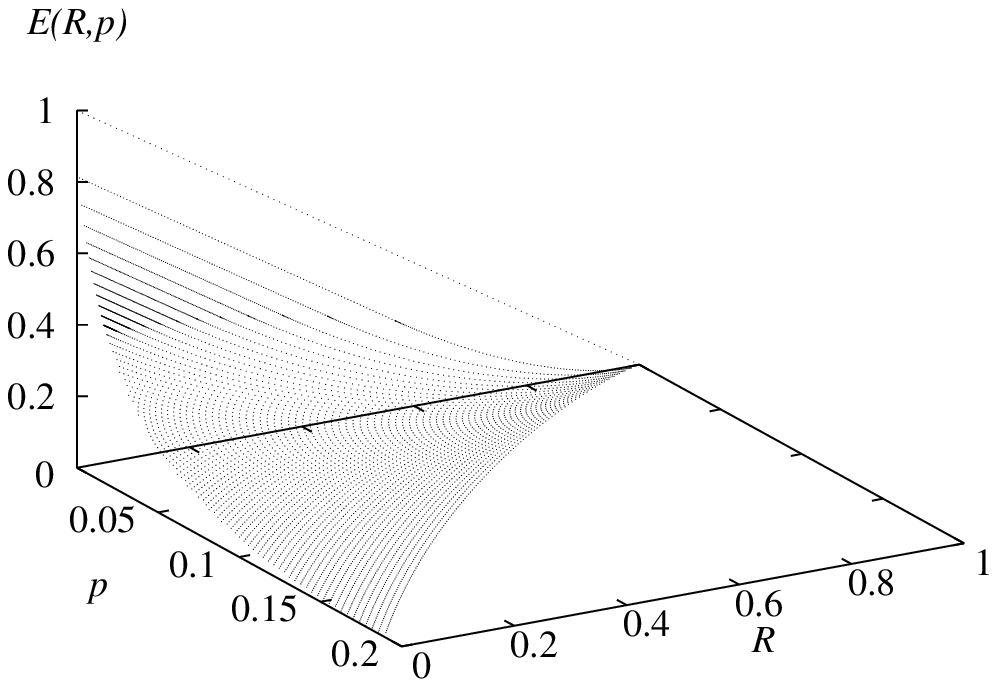}  
\end{center}
\caption{\label{fig:1} The function $E(R,P)=E(R,p)$ in the case where $\dmn=2$
and $P\big((0,0)\big)=1-p$, $P(u)=p/3$ for $u \ne (0,0)$, $u \in \cX = \{ 0,1 \}^2$,
which applies to the depolarizing channel.}
\end{figure}

Maximization of the bound $E(R,P_{\cU\cA,\Ebasis})$ or $1-H(P_{\cU\cA,\Ebasis})$
with respect to the basis $\Ebasis$ and the TPCP map $\cU$
seems troublesome and is largely left
untouched except for the following simple case. 
\begin{proposition}\label{prop:max}
Let a channel $\cA\sim\{A_x\}_{x\in\cX}$ be given
by $A_x=\sqrt{Q(x)}\, \tilde{\Ebe}_x$, $x\in\cX$,
where $\tilde{\Ebe}_{(i,j)}=\tilde{X}^i \tilde{Z}^j$, 
$\tilde{X}$ and $\tilde{Z}$ are defined by
\[
\tilde{X} |b_{j} \rag  = |b_{(j-1) \bmod \dmn } \rag, \quad
\tilde{Z} |b_{j} \rag = \tilde{\omega}^ j |b_{j} \rag
\]
similarly to (\ref{eq:error_basis}),
with $\{ \ket{b_{j}} \}$ and $\tilde{\omega}$ 
being an ONB of $\Hch$ and a primitive $\dmn$-th root of unity, respectively,
and $Q$ is a probability distribution on $\cX$.
Then,
the maximum of $1-H(P_{\cU\cA,\Ebasis})$ with respect to $\Ebasis$ and
$\cU$,
i.e., with respect to $\{ \ket{0},\dots\ket{\dmn-1}\}$, $\omega$
and $\cU$, where $\cU: \Bop(\Hch)\to \Bop(\Hch)$
ranges over all unitary maps,
is achieved by $\ket{j}=\ket{b_j}$, $j=0,\dots,\dmn-1$,
$\omega=\tilde{\omega}$, and $\cU=\Id$,
where $\Id$ denotes the identity map on $\Bop(\Hch)$.
\enprop

A proof is given in Appendix~A. 

Next, we consider general channels.
In a setting where elaborated coding schemes that rely on purification
protocols are allowed, the lower bound $1-H_1(p')$,
as well as the Shor-Smolin improvement on this,
for a general channel $\cA$ with $\dmn=2$
was known before this work~\cite{bennett96m,bennett96p}, \cite[the last paragraph]{ss97},
where 
\begin{equation}  \label{eq:pprime}
p'=
1-\max_{\eta} \bra{\eta}[\Id\tnsr \cA](\ket{\Phi^{+}}\bra{\Phi^{+}})
\ket{\eta},
\end{equation}
$\ket{\Phi^{+}}=2^{-1/2}(\ket{00} + \ket{11})$,
and the maximum is over all completely entangled states $\eta$.
We compare our bound with the bound $1-H_1(p')$,
which is `almost' the best among those previously
known in the sense
that the known improvement outperforms this only if
$1-0.8115=0.1885 \le p' \le 1-0.8094=0.1906$
and the difference between $1-H_1(p')$ and the improved one is
at most $10^{-2}$~\cite[Fig.~8]{dss98}. 
As is proved in Appendix~B, 
for every basis $\Ebasis$
defined with (\ref{eq:error_basis}) for some $\{ \ket{0},\ket{1}\}$,
where $\dmn=2$, 
there exists some unitary map $\cU$ satisfying
\begin{equation} \label{eq:bound_cmp}
1-H(P_{\cU\cA,\Ebasis}) \ge 1-H_1(p').
\end{equation}
Roughly speaking, the gain of this paper's bound comes from
the fact the bound has the form $1-H(P_{\cU\cA,\Ebasis})
=1-H\big((1-p',p_1,p_2,p_3)\big)=1-h(p')-p'H\big((p_1/p',p_2/p',p_3/p')\big)$,
and for a fixed $p'=p_1+p_2+p_3>0$, its minimum is $1-H_1(p')$
(reached when $p_1=p_2=p_3$); Bennett {\em et al.}\/'s
scheme~\cite{bennett96m} loses information on 
$M=[\Id\tnsr \cA](\ket{\Phi^{+}}\bra{\Phi^{+}})$
by `twirling' (a random bilateral rotation),
which increases entropy of $M$ as high as to $H_1(p')$.

The next example illustrates the advantage of this work.

{\em Example 1}.\/ 
Let us consider the amplitude-damping channel
whose Kraus operators are
\[
A_{(0,0)}=\begin{bmatrix} 1 & 0\\
0 & \sqrt{1-\gamma}   
\end{bmatrix} \quad \mbox{and} \quad
A_{(1,0)}=\begin{bmatrix} 0 & \sqrt{\gamma}   \\
0 & 0
\end{bmatrix}
\]
in matrix form with respect to the basis
$\{ \ket{0},\ket{1} \}$, 
where $\dmn=2$ and $0 \le \gamma \le 1$.
This channel has often been discussed as a reasonable
model in the context of quantum error correction%
~\cite[Section~3.4.2]{preskillLN}, \cite[Chapter~8]{nielsen_chuang}, \cite{KnillLaflamme97} while to this author's knowledge, it was not known if
any positive rates were achievable by
standard quantum error-correcting (stabilizer) codes on this channel.
The $A_{(0,0)}$ and $A_{(1,1)}$ can be expanded, respectively, as
\[
A_{(0,0)}=\frac{1+\sqrt{1-\gamma}}{2} I + 
\frac{1-\sqrt{1-\gamma}}{2} Z
\]
and
\[
 A_{(1,0)} = \frac{\sqrt{\gamma}}{2} (X - XZ).
\]
Regarding $A_{(0,1)}=A_{(1,1)}$ as the null operator,
we have 
\[
\begin{array}{ll}
 P_{\cA}\big((0,0)\big)=(2-\gamma+2\sqrt{1-\gamma})/4,  & 
 P_{\cA}\big((1,0)\big)=\gamma/4,\\
 P_{\cA}\big((0,1)\big)=(2-\gamma-2\sqrt{1-\gamma})/4,  &
 P_{\cA}\big((1,1)\big)=\gamma/4.
\end{array}
\]
Hence, our lower bound to the quantum capacity of this channel is 
\begin{equation}\label{eq:lb2amp}
1-H(P_{\cA})= 1-h\Big(\frac{\gamma}{2} \Big) - \Big(1-\frac{\gamma}{2}\Big)
 h\Bigg(\frac{1}{2}+\frac{\sqrt{1-\gamma}}{2-\gamma}\Bigg) -\frac{\gamma}{2}.
\end{equation}
This bound actually achieves the maximum of $1-H(P_{\cA,\Ebasis'})$
with respect to $\Ebasis'$
as can be checked by a direct calculation and the concavity of entropy.

This bound, together with
the previously known one $1-H_1(p')$ with (\ref{eq:pprime}),
is plotted in Fig.~\ref{fig:2}, where $p'$ is calculated in
Appendix~B, Example~4.
\begin{figure}
\begin{center}
\includegraphics[scale=1.0]{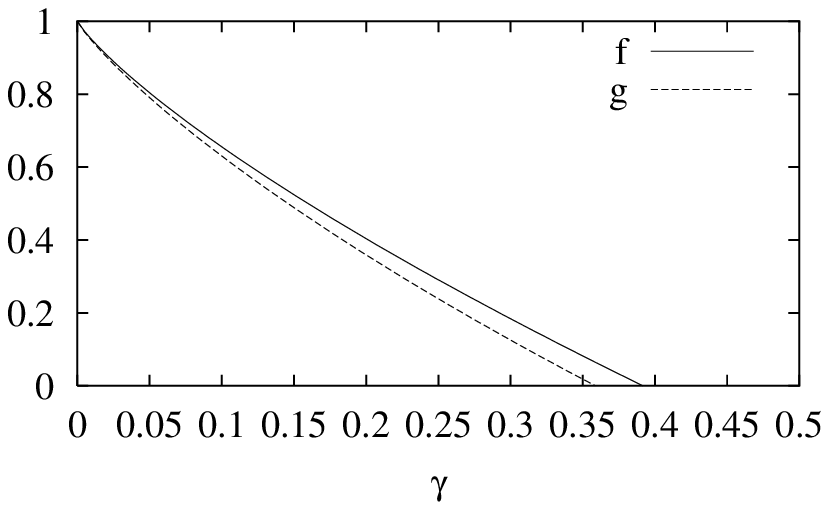}  
\end{center}
\caption{\label{fig:2} This paper's bound ${\rm f}=1-H(P_{\cA})$
in (\protect\ref{eq:lb2amp}), drawn as solid line,
and the previously known one ${\rm g}=1-H_1(p')$, dotted line,
with $p'=1-(2-\gamma+2\sqrt{1-\gamma})/4$
for the amplitude-damping channel in Example~1.
Shor and Smolin~\protect\cite{ss97,dss98} succeeded in improving
$1-H_1(p')$ by an amount 
less than $10^{-2}$ for some values of $p'$ with $1-H_1(p')<10^{-2}$.}
\end{figure}

%
\enexam

\section{Minimum Average Fidelity}

The minimum fidelity given in (\ref{eq:minimum_fidelity}) is the
simplest criterion for design of quantum error correction schemes.
A known substitute for the minimum fidelity is
the entanglement fidelity~\cite{schumacher96}.
It turns out that
yet another criterion is useful to establish Theorem~\ref{th:main}:
We seek codes of large {\em minimum average fidelity}.\/
The minimum average fidelity $\Fav(\Hcd)=\Fav(\Hcd,\cR \cA_n)$
of a code $(\Hcd,\cR)$ used on a channel 
$\cA_n  : \Bop(\Hch^{\tnsr n}) \to \Bop(\Hch^{\tnsr n})$
is defined by
\begin{equation} \label{eq:MAF}
\Fav(\Hcd)=
\min_{\Bcd}
\frac{1}{K} \sum_{\psi\in \Bcd} F(\psi, \cR \cA_n)
\end{equation}
where
$F(\psi, \cR \cA_n)= \lag\psi| \cR\cA_n(|\psi\rag \lag\psi|)
|\psi\rag$, $K$ is the dimension of $\Hcd$,
and the minimization with respect to $\Bcd$ is taken over all 
ONBs of $\Hcd$.
Note that the minimum exists since the minimization 
can be written as that of a
continuous function defined on a compact set.
According to Schumacher~\cite{schumacher96}, any average fidelity,
and hence the minimum average fidelity
are not less than the entanglement fidelity.

Employing the minimum average fidelity may need an account.
In the previous work~\cite{hamada01e}, 
Theorem~\ref{th:main} was proved for memoryless channels of the form
$\cA \sim \{ \sqrt{P(u)} \Ebe_{u} \}_{u\in\cX}$.
In this case,
$F(\Hcd)$ is trivially lower bounded by the sum of probabilities
of errors that are correctable by $\Hcd$.
The major difficulty in analysis on general channels
lies in the fact that
this bound is no longer true in general.
However, as we will see in the sequel,
a similar bound holds for a properly chosen symplectic quantum code
if we replace $F$ by the minimum average fidelity $\Fav$.
Furthermore, 
an estimate for $\Fav(\Hcd)$ automatically gives one for $F(\Hcd)$
by the following lemma.
\begin{lemma}\label{lem:av2min}
Let the minimum average fidelity
$\Fav(\Hcd)=\Fav(\Hcd,\cR \cA_n)$ of a code $(\Hcd,\cR)$
used on a channel
$\cA_n  : \Bop(\Hch^{\tnsr n}) \to \Bop(\Hch^{\tnsr n})$
satisfy
\[
 1-\Fav(\Hcd) \le \Pbad
\]
for some constant $\Pbad$, and assume $\Hcd$ has dimension $K \ge 2$.
Then, there exists a $\lfloor K/2 \rfloor$-dimensional subspace
$\cD$ of $\Hcd$ whose minimum fidelity 
$F(\cD)=F(\cD,\cR \cA_n)$
fulfills
\[
 1- F(\cD) \le 2 \Pbad.
\]
\mbox{}
\enlem

{\em Proof.}\/
Let a normalized vector $\psi_1$ minimize
$F(\psi) = \lag\psi| \cR\cA_n(|\psi\rag \lag\psi|)
|\psi\rag$ among those in $\Hcd$ ($=\Hcd_0$),
and let $\Hcd_1$ be the orthogonal complement of $\spn\{ \psi_1 \}$
in $\Hcd$,
which means $\Hcd=\Hcd_1 \oplus \spn\{ \psi_1 \}$.
Next, let $\psi_2$ minimize $F(\psi)$
among those in $\Hcd_1$, and 
let $\Hcd_2$ be the orthogonal complement of 
$\spn\{ \psi_1,\psi_2 \}$ in $\Hcd$,
which means $\Hcd=\Hcd_2 \oplus \spn\{ \psi_1,\psi_2 \}$.
Continue in the same way until we obtain $\psi_{\lceil K/2 \rceil}$
and $\Hcd_{\lceil K/2 \rceil}$. 
Put $\cD=\Hcd_{\lceil K/2 \rceil}$.
We annex an arbitrarily chosen ONB $\{ \psi_{\lceil K/2 \rceil+1}, \cdots,
\psi_{K} \}$ of $\cD$
to $\{ \psi_1,\dots,\psi_{\lceil K/2 \rceil} \}$
to form an ONB of $\Hcd$.
Now put $e(\psi)=1-F(\psi)$.
Then, by construction, 
\begin{eqnarray*}
1-F(\cD) & \le & e(\psi_{\lceil K/2 \rceil})\\
& \le & \frac{e(\psi_1)+ \dots + e(\psi_{\lceil K/2 \rceil})}{\lceil K/2 \rceil}\\
& \le & 2 \, \frac{e(\psi_1)+ \dots + e(\psi_{K})}{K}\\
& \le & 2 \Pbad,
\end{eqnarray*}
as promised.
\myQED

This lemma and its proof are analogous to those known 
in the classical information theory~\cite{gallager}, p.~140.
A similar idea was used by Barnum {\em et al.}~\cite{barnum00},
where they adopted entanglement fidelity in place of minimum average fidelity.
This lemma means that a properly chosen subcode $\cD$ of $\cC$
works without any loss of asymptotic performance.

\section{Codes based on Symplectic Geometry \label{ss:symplectic_codes}}

To prove the theorem, we use symplectic quantum codes, so that we
shall recall
basic facts on them in this section.
We can regard the index of $\Ebe_{(i,j)}=X^i Z^j$, $(i,j)\in\cX$,
as a pair of elements from the field $\myF=\field_{\dmn}=\bZ/\dmn\bZ$,
the finite field consisting of $\dmn$ elements.
From these, 
we obtain a basis $\Ebasis_n = \{ \Ebe_x \mid x \in (\myF^2)^n \}$
of $\Bop(\Hch^{\tnsr n})$,
where
$
\Ebe_x = \Ebe_{x_1} \tnsr \dots \tnsr \Ebe_{x_n}
$
for $x=(x_1,\dots,x_n)\in (\myF^2)^n$.
We write $\Ebe_{\Icr}$ for 
$\{ \Ebe_{x} \in \Ebasis_n \mid x\in\Icr \}$
where $\Icr \subset (\myF^2)^n$.
%
%
The index of a basis element
\[
\big((u_1,v_1),\dots,(u_n,v_n)\big)\in (\myF^2)^n
\]
can be regarded as the plain $2n$-dimensional vector
\[
x=(u_1,v_1,\dots,u_n,v_n) \in \myF^{2n}.
\]
We can equip the vector space $\myF^{2n}$ over $\myF$ with
a {\em symplectic bilinear form}\/ (symplectic pairing, 
or inner product), 
which is defined by
\begin{equation}\label{eq:symplectic_form}
\syp{x}{y} = \sum_{i=1}^{n} u_i v_i' - v_i u_i'
\end{equation}
for the above $x$ and $y=(u'_1,v'_1,\dots,u'_n,v'_n) \in \myF^{2n}$%
~\cite{artin,aschbacher}.
Given a subspace $\Cso \subset \myF^{2n}$, let
\[
\Cso^{\perp} = \{ x \in \myF^{2n} \mid \forall y\in \Cso,\ \syp{x}{y} =0 \}.
\]
\begin{lemma} \cite{crss97,crss98} 
\label{lem:symplectic_code} 
Let a subspace $\Cso\subset\myF^{2n}$ satisfy 
\begin{equation}\label{eq:self-orth}
\Cso\subset \Cso^{\perp} \quad \mbox{and} \quad \dim \Cso = n-\kpr.
\end{equation}
Choose a set $\Icr \subset \myF^{2n}$ such that
\begin{equation}\label{eq:J-correcting}
\{ y-x \mid x \in \Icr,\, y\in \Icr \} \subset (\Cso^{\perp} \setminus \Cso)\cmple,
\end{equation}
where the superscript ${\rm C}$ denotes complement.
Then, there exist $\dmn^{n-\kpr}$ subspaces of the form
\begin{equation}\label{eq:codespace}
\{ \psi \in \Hch^{\tnsr n} \mid \forall M\in\Ebe_{\Cso},\ M \psi =  \tau(M) \psi \}
\end{equation}
each of which has dimension $\dmn^\kpr$,
where $\tau(M)$ are scalars, and hence eigenvalues of $M\in \Ebe_{\Cso}$.
The direct sum of these subspaces is the whole space 
$\Hch^{\tnsr n}$ and each subspace together with a suitable
recovery operator serves as an $\Ebe_{\Icr}$-correcting quantum code.
\enlem

{\em Remarks.}\/
A precise definition of $\Ebe_{\Icr}$-correcting codes
can be found in 
Section~III of \cite{KnillLaflamme97}
and the above lemma 
has been verified with Theorem III.2 therein.
Most constructions of quantum error-correcting codes relies on this lemma,
which 
is valid
even if $\dmn$ is a prime other than two%
~\cite{knill96a,knill96b,rains99}; 
related topics have been discussed in \cite{Gottesman99,MatsumotoUyematsu00,AshikhminKnill00}.
In this paper, we call the quantum codes in Lemma~\ref{lem:symplectic_code}
symplectic quantum codes or
{\em symplectic codes}\/ while Rains~\cite{rains99} indicates $\Cso$
by the latter term.
Symplectic codes are often called
additive codes~\cite{crss97,crss98} or 
stabilizer codes~\cite{gottesman96,gottesmanPhD},
and the set $\Ebe_{\Cso}$ in the lemma is called a stabilizer
in the literature.
\enremark

The next lemma, which immediately follows from
Lemma~\ref{lem:symplectic_code}, will be used in the proof of
 Theorem~\ref{th:main} below.
\begin{lemma} \cite{crss97,crss98} \label{lem:coset_leaders}
As in Lemma~\ref{lem:symplectic_code}, 
assume a subspace $\Cso\subset\myF^{2n}$ 
satisfies (\ref{eq:self-orth}).
In addition, let $\Icr_0 \subset \myF^{2n}$ 
be a set satisfying
\begin{equation}\label{eq:coset_leaders}
\forall x,y\in \Icr_0,\ [ \, y-x \in \Cso^{\perp} \Rightarrow x=y \, ].
\end{equation}
Then, the condition (\ref{eq:J-correcting}) is fulfilled,
so that the $\dmn^{n-\kpr}$ codes of the form (\ref{eq:codespace})
are $\dmn^{\kpr}$-dimensional $\Ebe_{\Icr_0}$-correcting codes.
\enlem

We assume the next in what follows.

{\em Assumption.}\/ 
When we speak of an $\Ebe_{\Icr}$-correcting symplectic code $\Hcd$,
the recovery operator $\cR$ for 
the code is always the one presented by Knill and Laflamme%
~\cite{KnillLaflamme97}, proof of Theorem III.2.
\hfill $\Diamond$ 

Note that the $\cR$ is determined from $\Hcd$ and $\Icr$ in general.
In the present case where $\Hcd$ is a symplectic quantum code
in Lemma~\ref{lem:coset_leaders} 
(or Lemma~\ref{lem:coset_leaders_refined} below),
the recovery operator $\cR$ can be written explicitly,
viz., $\cR \sim \{ \Pi_{\rm rest} \} \cup \{ \Ebe_{r}^{\dagger} \Pi_r
\}_{r\in\Icr_0}$,
where $\Pi_r$ is the projection onto
$\Ebe_r \Hcd= \{ \Ebe_r\psi \mid \psi\in\Hcd \}$,
and $\Pi_{\rm rest}$ is the projection onto the orthogonal complement
of $\bigoplus_{r\in\Icr_0} \Ebe_r \Hcd$ in $\Hch^{\tnsr n}$.
The premise (\ref{eq:coset_leaders}) of Lemma~\ref{lem:coset_leaders}
can be restated as that $\Icr_0$ is a set of representatives of cosets
of $\Cso^{\perp}$ in $\myF^{2n}$.
When the code is used on a channel
$\cA_{n}\sim \{ \sqrt{P_n(x)} \Ebe_{x} \}$,
a natural choice for $\Icr_0$ would be a set
consisting of representatives each of which maximizes
the probability $P_{n}(x)$ in the coset%
~\cite{crss98}
since it is analogous to maximum likelihood decoding,
which is an optimum strategy for classical coding
(see Slepian~\cite{slepian56} or any textbook of information theory). 
In the proof below, 
we choose another set of representatives,
the classical counterpart of which 
(minimum entropy decoding) asymptotically yields the same performance
as maximum likelihood decoding~\cite{csiszar_koerner,CsiszarKoerner81a}.

\section{Bound on Minimum Average Fidelity}

\subsection{Plan of Proof \label{ss:plan}}

Our strategy for proving Theorem~\ref{th:main} is to employ
the random coding technique known in classical information theory%
~\cite{gallager65,gallager,goppa74,csiszar_koerner}.
A typical random coding argument goes as follows.
Suppose $F'(\cC)$ is a measure
of performance, which is the minimum average fidelity in our case, 
of a code $\cC$ and
we want to prove the existence of a code $\cC$ with $F'(\cC)\ge G$.
We take some ensemble $\cE$ of codes,
and evaluate the ensemble average $\crd{\cE}^{-1}\sum_{\cC\in\cE} F'(\cC)$.
If the average is lower bounded by $G$, then we can conclude
at least one code $\cC$ in $\cE$ has performance not smaller than $G$.
In what follows, we will use this proof method twice, that is,
first, with $\Cso$ fixed and
$\cE$ being the set, say $\cE(\Cso)$,
of $\dmn^{n-m}$ subspaces in Lemma~\ref{lem:symplectic_code} or \ref{lem:coset_leaders},
and second, with $\cE$ consisting of all $\Cso$ 
satisfying (\ref{eq:self-orth}).

\subsection{Preskill's Lower Bound on Fidelity}

Preskill showed an interesting lower bounds on the minimum fidelity
of a code used on quantum channels,
which will be presented in a slightly different form here.
\begin{lemma}
\cite{preskillLN} \label{lem:Preskill_bound}
For a channel
$\cA_n  : \Bop(\Hch^{\tnsr n}) \to \Bop(\Hch^{\tnsr n})$,
an $\Ebe_{\Icr}$-correcting code $(\Hcd \subset \Hch^{\tnsr n}, \cR)$
and any state $\ket{\psi} \in\Hcd$,
the fidelity 
$F(\psi)=\lag\psi| \cR\cA_n(|\psi\rag \lag\psi|) |\psi\rag$
is bounded by
\[
F(\psi) \ge 1 - \sum_{x\in\cX^n} \bra{\psi} B_x^{\dagger} B_x \ket{\psi}
\]
where $B_x=\sum_{y\in \Icr\cmple} a_{xy}\Ebe_y$, $x\in \cX^n$.
\enlem

This is Preskill's lower bound%
~\cite{preskillLN}, Section~7.4.1, Eq.~(7.58),
and the above form can be obtained by
rewriting the channel, which was described in terms of
unitary evolution of a state of an enlarged system
and a partial trace operation,
into an operator-sum representation.
In Appendix~C,
an alternative proof which uses only operator-sum representations
is presented.

\subsection{Minimum Average Fidelity Bound for Symplectic Codes}


To evaluate the minimum average fidelity of codes,
we first associate a sequence of probability distributions $\{ P_{\cA_n} \}$
with the channel $\{ \cA_n \}$ on which codes are to be evaluated.
\begin{definition}\label{def:P4An}
For each $n$, let $\cA_n \sim \{ A_x^{(n)} \}_{x\in\cX^n}$,
expand $A_x^{(n)}$ as 
$A_x^{(n)} = \sum_{y\in\cX^n} a_{xy} \Ebe_y$, $x\in\cX^n$,
and define a probability distribution $P_{\cA_n}$ on $\cX^n$ by 
\[
P_{\cA_n}(y) = \sum_{x} |a_{xy}|^2, \quad y\in\cX^n.
\]
\mbox{}
\closedef

That $\sum_{x\in\cX^n} P_{\cA_n}(x)=1$ readily follows, again,
from the trace-preserving condition
$\sum_{x\in\cX^n} A_x^{(n)\dagger} A_x^{(n)} =I$ and the property of the basis $\Ebasis_n$
that $\Ebe_x^{\dagger} \Ebe_y = I$ if and only if $x=y$~\cite{knill96a}.

{\em Example 2.}\/
Let $\{\cA_n\}$ be a memoryless channel
$\cA_n=\cA^{\tnsr n}, n=1,2,\dots$. 
It is easy to see that
\begin{equation}\label{eq:memorylessP}
P_{\cA_n}(y_1,\dots,y_n) = \prod_{i=1}^{n} P_{\cA}(y_i)
\end{equation}
where $P_{\cA}=P$ has already appeared in Definition~\ref{def:P4A}.
\enexam

The next is a result of the first application of random coding technique
in this paper.
\begin{lemma}\label{lem:fidelity_symplectic_codes}
As in Lemma~\ref{lem:symplectic_code}, 
let a subspace $\Cso\subset\myF^{2n}$ satisfy 
(\ref{eq:self-orth}) and (\ref{eq:J-correcting})
with some $\Icr\subset \myF^{2n}$,
and let
$\cA_n  : \Bop(\Hch^{\tnsr n}) \to \Bop(\Hch^{\tnsr n})$
be a channel (TPCP linear map). With $\Cso$, $\Icr$ and $\cA_n$ fixed,
let $\Hcd(\Cso)$ achieve
the maximum of $\Fav(\Hcd)=\Fav(\Hcd,\cR\cA_n)$ 
in $\cE(\Cso)$ (see Section~\ref{ss:plan}), i.e., 
the maximum among the $\dmn^{n-\kpr}$ symplectic
codes associated with $L$ as in Lemma~\ref{lem:symplectic_code} or \ref{lem:coset_leaders}.
Then,
\[
1-\Fav\big(\Hcd(\Cso)\big) \le \sum_{x\notin {\Icr}} P_{\cA_n}(x).
\]
\enlem

{\em Proof.}\/
Taking the averages over an ONB $\Bcd$ of a code $\Hcd$
of both sides of the inequality 
in Lemma~\ref{lem:Preskill_bound}, we have
\[
1 - \frac{1}{\dmn^{\kpr}} \sum_{\psi\in\Bcd} F(\psi) 
\le \frac{1}{\dmn^{\kpr}} \sum_{\psi\in\Bcd} 
\sum_{x} \bra{\psi} B_x^{\dagger}B_x \ket{\psi},
\]
This holds for all ONBs $\Bcd$ of $\Hcd$
including the worst one $\Bcd_{\star}(\Hcd)$, which is a minimizer
for (\ref{eq:MAF}), so that
\[
1-\Fav(\Hcd) \le \frac{1}{\dmn^{\kpr}} \sum_{\psi\in\Bcd_{\star}(\Hcd)} 
\sum_{x} \bra{\psi} B_x^{\dagger}B_x \ket{\psi}.
\]
With $\Cso$ fixed, we have $\dmn^{n-\kpr}$ choices for $\Hcd$.
Taking the averages of both sides of the above inequality over these choices,
we obtain
\begin{eqnarray*}
\lefteqn{\frac{1}{\dmn^{n-\kpr}} \sum_{\Hcd} [1-\Fav(\Hcd)]}&&\\
& \le &
\frac{1}{\dmn^{n-\kpr}} \sum_{\Hcd} 
\frac{1}{\dmn^{\kpr}} \sum_{\psi\in\Bcd_{\star}(\Hcd)} 
\sum_{x} \bra{\psi} B_x^{\dagger} B_x \ket{\psi} \\
& = &
\frac{1}{\dmn^n} \sum_{x} \sum_{\Hcd} \sum_{\psi\in\Bcd_{\star}(\Hcd)} 
\bra{\psi} B_x^{\dagger} B_x \ket{\psi}\\
&= & \frac{1}{\dmn^{n}} \sum_{x} {\rm Tr} B_x^{\dagger} B_x\\
&= & \frac{1}{\dmn^{n}} \sum_{x} {\rm Tr} \sum_{y,z\in\Icr\cmple} a_{xy}^* \Ebe_{y}^{\dagger} a_{xz} \Ebe_{z} \\
&=& \sum_{x} 
\sum_{y\in\Icr\cmple} |a_{xy}|^2\\
&=& \sum_{y\in J\cmple} P_{\cA_n}(y),
\end{eqnarray*}
where we have used the fact
that the $\dmn^{n-m}$ subspaces $\Hcd$ sum to
$\Hch^{\tnsr n}$ orthogonally for 
the second equality, 
and the property of error basis $\Ebasis_n$ that
${\rm Tr} \Ebe_y^{\dagger} \Ebe_z = \dmn^n \delta_{yz}$
for the fourth equality~\cite{knill96a}.
Hence, at least, one code $(\Hcd,\cR)$ has the promised minimum average 
fidelity. \myQED

{\em Example 3.}\/ 
To illustrate the difference between the minimum average fidelity $\Fav$
and minimum fidelity $F$ as well as the 
significance of Lemma~\ref{lem:fidelity_symplectic_codes},
let us consider again the amplitude-damping channel discussed in Example~1
and evaluate some small codes on this channel.
Let $n=2$ and $\kpr=1$.
In this example, we denote
a vector $(u_1,v_1,u_2,v_2)\in\myF^4$ simply by $u_1v_1u_2v_2$.
Let $\Cso= \{ 0000, 0101 \}$.
Then, $\Ebe_{\Cso}=\{ I \tnsr I, Z\tnsr Z \}$,
and we have two symplectic codes $\Hcd_0 = \spn\{ \ket{00}, \ket{11} \}$
and $\Hcd_1 = \spn\{ \ket{01}, \ket{10} \}$, where $\ket{00}
=\ket{0}\tnsr\ket{0}$ and so on.
It is easy to check that the cosets of $\Cso^{\perp}$ in $\myF^{4}$ are
\[
\Cso^{\perp} \ = \  \{ 0000,0101,1010,1111,0001,0100,1011,1110 \}
\]
and
\[
h_1+\Cso^{\perp} =  \{ 1000,1101,0010,0111,1001,1100,0011,0110 \},
\]
where $h_1=1000$.
Let $\Pi_0$ and $\Pi_{1}$ denote the projections onto
$\Hcd_0$ and $\Hcd_1$, respectively.
Putting $\Icr_{0} = \{ 0000, h_1 \}$
and $\Icr=h_1+\Cso=\{ 0000,0101,1000,1101 \}$, we see that
both $(\Hcd_0,\cR_0)$ and $(\Hcd_1,\cR_1)$, where
$\cR_0 \sim \{ \Pi_0, \Ebe_{h_1}^{\dagger} \Pi_{1} \}$
and $\cR_1 \sim \{ \Pi_{1}, \Ebe_{h_1}^{\dagger} \Pi_0 \}$,
are $\Ebe_{\Icr_0}$-correcting as well as $\Ebe_{\Icr}$-correcting
from Lemmas~\ref{lem:coset_leaders} and \ref{lem:coset_leaders_refined}
or directly from Lemma~\ref{lem:symplectic_code}
(recall also the general form of $\cR$ for a symplectic code
was given in the 
the last paragraph of
Section~\ref{ss:symplectic_codes}).
If we prepare an input state $\ket{\psi}=x\ket{00}+y\ket{11}\in\Hcd_0$,
then, the fidelity
$F(\psi)=\bra{\psi} \cR_0 \cA^{\tnsr 2} (\ket{\psi}\bra{\psi}) \ket{\psi}$
can be calculated as $1 - \gamma y y^*$.
This implies the minimum fidelity is $F(\Hcd_0)=1-\gamma$
while the minimum average fidelity is $\Fav(\Hcd_0)=1-\gamma/2$.
In a similar way, evaluating $F(\Hcd_1)$ results in $F(\Hcd_1)=1-\gamma$
and $\Fav(\Hcd_1)=1-\gamma/2$.
One the other hand, the bound in 
Lemma~\ref{lem:fidelity_symplectic_codes} states
$\Fav\big(\Hcd(\Cso)\big) \ge 1-\sum_{z\notin\Icr} 
P_{\cA^{\tnsr 2}}(z)= \sum_{z\in\Icr}
P_{\cA}^{2}(z)=1-3\gamma/4$, where  $P_{\cA}^{2}$ is the product
measure obtained from $P_{\cA}$ as in (\ref{eq:memorylessP}).
This is an example for which the inequality in 
Lemma~\ref{lem:fidelity_symplectic_codes}
is true but that with $\Fav$ replaced by 
$F$ fails.
\enexam

\section{Proof of Theorem~\protect\ref{th:main}}


We put $P=P_{\cA,\Ebasis}$.
Since the bound in the theorem is trivial when
$k \ge n-1$, we assume $\kpr=k+1 < n$.
What we want is a code $(\cD,\cR)$ with dimension $\dmn^k$
whose minimum fidelity is lower bounded by
$1 - 2 \dmn^2 (n+1)^{2(\dmn^2-1)} \dmn ^{ - n E(R,P) }$.
To show the existence of such a code, it is enough to prove 
\begin{equation}\label{eq:goal_av}
1-\Fav\big(\Hcd(\Cso)\big) \le \dmn^2 (n+1)^{2(\dmn^2-1)} \dmn ^{ - n E(R,P) }
\end{equation}
for some $\Cso$ with $\dim \Cso = n-\kpr = n-(k+1)$
and some choice of $\Icr_0$ in Lemma~\ref{lem:coset_leaders},
where $\Hcd(\Cso)$ achieves
the maximum of $\Fav(\Hcd)=\Fav(\Hcd,\cR\cA)$
among the $\dmn^{n-\kpr}$ symplectic
codes associated with $L$ as in
Lemma~\ref{lem:fidelity_symplectic_codes}, 
since we have Lemma~\ref{lem:av2min}.
Recall that the probability distribution $P_{\cA_n}$
for the memoryless channel $\cA$ has a product
form as in (\ref{eq:memorylessP}), which is denoted by $P^n$
in this proof.

We employ the method of types%
~\cite{csiszar_koerner,csiszar98,CsiszarKoerner81a,cover_th},
on which a few basic facts to be used are 
collected here.
For $x=(x_1,\dots,x_n)\in\cX^n$,
define a probability distribution $\sP_{x}$
on $\cX$ by 
\[
\sP_{x}(u)=\frac{\crd{\{ i \mid 1\le i \le n, x_i = u \}}}{n}, \quad u \in \cX,
\]
which is called the {\em type}\/ (empirical distribution) of $x$. 
With $\cX$ fixed, the set of all possible types of sequences from
$\cX^n$ is denoted by $\cQ_n(\cX)$ or simply by $\cQ_n$.
For a type $Q\in \cQ_n$, $\cT_{Q}^n$ is defined as
$\{ x\in\cX^n \mid \sP_{x} = Q \}$.
In what follows, we use 
\begin{equation}\label{eq:types1}
\crd{\cQ_n} \le (n+1)^{\crd{\cX}-1} ,
\end{equation}
where $\crd{\cX}=\dmn^2$ in the present case, and
\begin{equation}\label{eq:types2}
\forall Q\in \cQ_n,\quad
|\cT_{Q}^n| \le \dmn^{nH(Q)}.
\end{equation}
Note that if $x\in\cX^{n}$ has type $Q$, then
$P^{n}(x)=\prod_{a\in\cX} P(a)^{nQ(a)} = \exp_{\dmn} \{ -n [H(Q)+D(Q||P)]
\}$. 

We apply Lemma~\ref{lem:coset_leaders}
choosing $\Icr_0$ as follows.
Since $\dim \Cso = n-\kpr$,
we have $\dim \Cso^{\perp} = n+\kpr$~\cite{artin,grove}.
From each of the $\dmn^{n-\kpr}$ cosets of $\Cso^{\perp}$ in $\myF^{2n}$,
select a vector that minimizes $H(\sP_{x})$, i.e., a vector $x$ satisfying
$H(\sP_{x})\le H(\sP_{y})$ for any $y$ in the coset.
Let $\Icr_0(\Cso)$ denote the set of the $\dmn^{n-\kpr}$ selected vectors.
This selection uses the idea of the minimum entropy 
decoder known in the classical information theory literature%
~\cite{CsiszarKoerner81a}.
Let
\[
\Aso = \{ \Cso \subset \myF^{2n} \mid \mbox{$\Cso$ linear}, \ \Cso \subset
\Cso^{\perp},\ \dim \Cso = n-\kpr \}
\]
and  for each $L\in\Aso$, 
let $\Hcd(\Cso)$ be the best $\Ebe_{\Icr_0(\Cso)}$-correcting code
in $\cE(\Cso)$. 
Putting
\[
\Fbar = \frac{1}{\crd{\Aso}} \sum_{\Cso\in\Aso} \Fav\big(\Hcd(\Cso)\big),
\]
we will show that $1-\Fbar$ is bounded from above by
$\dmn^2 (n+1)^{2(\dmn^2-1)} \dmn ^{ - n E(R,P) }$,
which will ensure (\ref{eq:goal_av}) for some $\Cso$ and hence,
establish the theorem by the 
argument at the beginning of this proof.
This is our second application of the random coding method.

The $\{0,1\}$-valued
indicator function 
$\indc [ T ]$ equals 1 if the statement $T$ is true and
equals 0 otherwise.
From Lemma~\ref{lem:fidelity_symplectic_codes}, we have
\begin{eqnarray}
1- \Fbar & \le & \frac{1}{\crd{\Aso}} \sum_{\Cso\in\Aso} \sum_{x \notin \Icr_0(\Cso)}  P^n(x) \nonumber\\
     & = & \frac{1}{\crd{\Aso}} \sum_{\Cso\in\Aso} \sum_{x \in \myF^{2n}}  
             P^n(x) \indc [ x \notin \Icr_0(\Cso) ]  \nonumber\\
     & = & \sum_{x \in \myF^{2n}} P^n(x) \frac{\crd{\Bcn(x)}}{\crd{\Aso}},
\label{eq:pr0}
\end{eqnarray}
where we have put 
\[
\Bcn(x) = \{ \Cso \in \Aso \mid x \notin \Icr_0(\Cso) \}, \quad x\in\myF^{2n}.
\]
The fraction $\crd{\Bcn(x)}/\crd{\Aso}$ is trivially bounded as
\begin{equation} \label{eq:pr1}
\frac{\crd{\Bcn(x)}}{\crd{\Aso}} \le 1, \quad x\in\myF^{2n}.
\end{equation}
We use the next lemma, a proof of which is given in Appendix~D. 
\begin{lemma}\label{lem:C_uniform}
Let
\[
\Acn(x) = \{ \Cso \in \Aso \mid x \in \Cso^{\perp} \setminus \{ 0 \} \}.
\]
Then, $\crd{\Acn(0)}=0$ and
\begin{equation}\label{eq:C_uniform}
\frac{\crd{\Acn(x)}}{\crd{\Aso}}
 = \frac{\dmn^{n+\kpr} - 1}{\dmn^{2n}-1}  
\le \frac{1}{\dmn^{n-\kpr}}, \quad x\in\myF^{2n},\
 x \ne 0.
\end{equation}
\enlem

{\em Remarks.}\/ 
Note that $\Acn$ is not empty 
since any $(n-m)$-dimensional subspace of
\[
 \{ (x_1,0,x_3,0,\dots,x_{2n-1},0) \in \myF^{2n} 
\mid x_1,x_3,\dots,x_{2n-1} \in \myF \}
\]
is contained in $\Acn$.
This lemma is essentially 
due to Calderbank {\em et al.}~\cite{crss97} 
who have used it with $\Aso(x)$ replaced by
$\{ \Cso \in \Aso' \mid x \in \Cso^{\perp} \setminus \Cso \}$
for some $\Aso'\subset\Aso$
to prove the Gilbert-Varshamov-type bound for quantum codes.
Matsumoto and Uyematsu~\cite{MatsumotoUyematsu01}
proved Lemma~\ref{lem:C_uniform}
with $\Aso(x)$ replaced by
$\{ \Cso \in \Aso \mid x \in \Cso^{\perp} \setminus \Cso \}$
using the Witt lemma explicitly~\cite{artin,aschbacher}.
The present definition of $\Aso(x)$ makes the argument easier.
\enremark

Since $\Bcn(x) \subset \{ \Cso \in \Aso \mid \exists y\in\myF^{2n}, H(\sP_y) \le H(\sP_x), y-x\in \Cso^{\perp} \setminus \{ 0 \} \}$
from the design of $\Icr_0(\Cso)$ specified above (cf.~\cite{goppa74}), 
\begin{eqnarray}
 \crd{\Bcn(x)} &\le &\sum_{y\in \myF^{2n} :\, H(\sP_y) \le  H(\sP_x),\ y \ne x}
 \crd{\Acn(y-x)}\nonumber\\
  &\le & \sum_{y\in \myF^{2n} :\, H(\sP_y) \le  H(\sP_x),\ y \ne x}
\crd{\Aso}{\dmn}^{-n+\kpr}, \label{eq:pr2} 
\end{eqnarray}
where we have used (\ref{eq:C_uniform}) for the latter inequality.
Combining (\ref{eq:pr0}), (\ref{eq:pr1}) and (\ref{eq:pr2}), we can proceed
as follows with the aid of the basic inequalities in (\ref{eq:types1})
and (\ref{eq:types2}) as well as the inequality
$\min \{ a+b, 1\} \le \min \{ a, 1\} + \min \{ b, 1\}$ for $a,b \ge 0$:
\begin{eqnarray*}
1-\Fbar &\le& \sum_{x\in\myF^{2n}} P^n(x) \ \min \Biggl\{ \ \sum_{y\in\myF^{2n} :\, H(\sP_y) \le  H(\sP_x),\ y \ne x} \dmn^{-(n-\kpr)},\ 1 \ \Biggr\}\\
 & \le & \sum_{Q\in \cQ_n}  \crd{\cT_{Q}^n} \prod_{a\in\cX} P(a)^{nQ(a)}
\ \min \Biggr\{ \sum_{Q'\in \cQ_n :\, H(Q') \le H(Q)} \frac{ \dmn^2 |\cT_{Q'}^n|}{\dmn^{n(1-R)}}, \ 1\ \Biggl\}\\
& \le & \dmn^2 \sum_{Q\in\cQ_n} \exp_{\dmn} [ -n D(Q || P) ] 
\sum_{Q'\in \cQ_n :\, H(Q')\le H(Q)} \exp_{\dmn} [ -n |1-R-H(Q')|^{+} ]\\
& \le & \dmn^2 \sum_{Q\in\cQ_n} \exp_{\dmn} [ -n D(Q || P) ] \,
|\cQ_n| \max_{Q'\in\cQ_n :\, H(Q') \le H(Q)}\exp_{\dmn} [ -n |1-R-H(Q')|^{+} ]\\
& = & \dmn^2 \sum_{Q\in\cQ_n} |\cQ_n| \exp_{\dmn} [ -n D(Q||P) -n|1-R-H(Q)|^+ ]\\
& \le & \dmn^2 |\cQ_n|^2 \exp_{\dmn} \big\{ \max_{Q\in\cQ_n} 
[ -n D(Q||P) -n|1-R-H(Q)|^+ ] \big\}\\
&\le &  \dmn^2 (n+1)^{2(\dmn^2-1)} \exp_{\dmn} [ -n E(R,P) ].
\end{eqnarray*}
This implies at least one $\Cso$ satisfies (\ref{eq:goal_av}),
and the proof is complete owing to Lemma~\ref{lem:av2min}.
\myQED

\section{Concluding Remarks}

This paper provided evidence,
from an information theoretic viewpoint,
that standard quantum error correction
schemes work reliably in the presence of quantum noise,
the effects of which are modeled as general completely positive linear maps.
What is technically new is evaluating the minimum average fidelity
over all eigenspaces of a stabilizer $\Ebe_{\Cso}$,
which yields a good estimate for the minimum fidelity of codes.
The thus obtained fact (Lemma~\ref{lem:fidelity_symplectic_codes})
allowed us to derive the main result in a manner familiar
in information theory.
Likewise, based on Lemma~\ref{lem:fidelity_symplectic_codes}
and with another classical technique, a high-rate improvement,
which corresponds to the expurgated bound in classical channel coding,
on the exponent $E(R,P)$
has already made in \cite{barg02} after the online distribution of
the present work,
though it is effective only for channels of low noise level 
and does not improve the capacity bound.

Although this paper's lower bound on the capacity is the best among those known
except for a few cases,
it is important to recognize that this paper's lower bound is not tight
in general. In this sense, Shor and Smolin~\cite{ss97,dss98} have gone
further.
Specifically, Shor and Smolin exploited the `degeneracy' of 
error-correcting codes
to present a lower bound on the capacity 
of the depolarizing channel
$\cA \sim \{\sqrt{1-p} \, I,  \sqrt{p/3}\, X,
\sqrt{p/3}\, XZ, \sqrt{p/3}\, Z \}$
such that their bound is positive while the bound
$1-H(P_{\cA})= 1- h(p) - p \log_2 3$
becomes negative for restricted values of $p$, 
where $h$ is the binary entropy function.
The degeneracy concept
is somewhat misleading because a single quantum code
can be regarded as both degenerate and nondegenerate
as is clearly understood from the next lemma, 
which is a refinement of 
Lemma~\ref{lem:symplectic_code}.
\begin{lemma}\label{lem:coset_leaders_refined}
As in Lemma~\ref{lem:coset_leaders},
assume a subspace $\Cso\subset\myF^{2n}$ and $\Icr_0$ 
satisfy (\ref{eq:self-orth}) and (\ref{eq:coset_leaders}), respectively.
Put
\[
 \Icr=\{ z+w \mid z\in\Icr_0,w\in\Cso \}.
\]
Then, the condition (\ref{eq:J-correcting}) is fulfilled,
so that the $\dmn^{n-\kpr}$ codes of the form (\ref{eq:codespace})
are $\dmn^{\kpr}$-dimensional $\Ebe_\Icr$-correcting codes.
\enlem

If an $\Ebe_{J'}$-correcting code is given and
$\{ M \ket{\psi} \mid M \in \Ebe_{\Icr'} \}$ 
is not linearly independent for a state $\ket{\psi}$ in the code space, 
then the code is called degenerate~\cite{crss98}.
The codes in Lemma~\ref{lem:coset_leaders_refined}
are nondegenerate $\Ebe_{\Icr_0}$-correcting codes
while they are degenerate $\Ebe_{\Icr}$-correcting codes.
In this paper, we have evaluated
nondegenerate $\Ebe_{\Icr_0}$-correcting codes
with $\crd{\Icr_0}=\dmn^{n-\kpr}$, but actually
$\crd{\Icr}=\dmn^{2(n-\kpr)}$ in this case.
Hence, the codes can correct more errors than those
evaluated in this paper.
Suggestions for developing Shor and Smolin's result can
be found in the final section of \cite{dss98}.

Shor and Smolin's result
does not deny the possibility of the tightness of this paper's bound
for all channels.
Extending this work's result to the case of
channels with memory of a Markovian nature is possible if
second-order (or higher-order) types are used instead
of the usual types~\cite{hamada02m}.
It may be also interesting to ask whether the present approach
will help us obtain bounds or improve the known ones
for Gaussian quantum channels already discussed
in the literature~\cite{HolevoWerner01,GottesmanKP01,HPreskill01}.

\section*{Acknowledgment}

The author 
wishes to thank R.~Matsumoto for 
valuable discussions and comments,
especially, for drawing this author's attention to
the problem of lower bounding the quantum capacity of general
memoryless channels using standard quantum codes,
M.~Hayashi and A.~Barg for helpful comments,
H.~Imai and K.~Matsumoto for support. 

\appendices

\subsection{Proof of Proposition~\protect\ref{prop:max} \label{app:1}}

In this proof, we assume
$\dmn=2$ for notational simplicity.
The proof readily extends to the case where $\dmn>2$.
First, we show that the maximum of $1-H(P_{\cU\cA,\Ebasis})$
with the restriction $\cU=\Id$
is achieved by the indicated $\Ebasis$.
For $\cM: \Bop(\Hch^{\tnsr 2}) \to \Bop(\Hch^{\tnsr 2})$
and $4 \times 4$ matrices $M$ over $\bC$,
we write $\cM \mateq M$ if $M$ is the
matrix of $\cM$ 
with respect to the basis $\{ \ket{b_0b_0}, \ket{b_0b_1}, \ket{b_1b_0}, \ket{b_1b_1} \}$,
where $\ket{b_0b_0}=\ket{b_0}\tnsr\ket{b_0}$ and so on.
We use the next lemma due to Choi~\cite{choi75,RuskaiSW01}.

\begin{lemma} \label{lem:choi} \cite{choi75}
A linear map $\cA: \Bop(\Hch) \to \Bop(\Hch)$
is completely positive if and only if
$[\Id\tnsr \cA]( \ket{\Phi^{+}} \bra{\Phi^{+}} ) $
is positive, 
where $\Id$ is the identity map on $\Bop(\Hch)$, and
\[
\ket{\Phi^+} = \frac{1}{\sqrt{2}}(\ket{b_0b_0} + \ket{b_1b_1}).
\]
Moreover, 
if we represent $[\Id\tnsr \cA]( \ket{\Phi^{+}} \bra{\Phi^{+}} )$ as
\begin{equation}\label{eq:chois_matrix2}
[\Id\tnsr \cA]( \ket{\Phi^{+}} \bra{\Phi^{+}} ) 
\mateq \frac{1}{2} \sum_{x\in\cX} \mbm{a}_{x}^{\dagger} \mbm{a}_{x}
\end{equation}
and rearrange the elements of 
$\mbm{a}_{x}=(\ahat_{x,00},\ahat_{x,01},\ahat_{x,10},\ahat_{x,11}) \in 
\bC^{4}$
into the matrix form 
\[
 \hat{A}_{x}= \begin{bmatrix}\ahat_{x,00} & \ahat_{x,01}\\
\ahat_{x,10} & \ahat_{x,11}
\end{bmatrix},\quad x\in\cX,
\]
we obtain an operator-sum representation of $\cA$:
$\cA \sim \{ A_{x} \}$,
where $A_{x}: \Bop(\Hch) \to \Bop(\Hch)$
is the Hermitian adjoint operator of
$\sum_{(i,j)\in\cX} \ahat_{x,ij} \ket{b_i}\bra{b_j}$,
i.e., the adjoint of the operator whose matrix
is $\hat{A}_x$, $x\in\cX$.
\enlem

{\em Remark}.\/ The correspondence $\xi:\, \bC^{4} \to \Bop(\Hch)$ 
that has sent $\mbm{a}_{x}$ to $A_x$ 
is explicitly written as 
\[
\xi(m_{00},m_{01},m_{10},m_{11})
= \sum_{(i,j)\in\cX} m_{ij}^* \ket{b_j} \bra{b_i}.
\] 
\enremark

If we define an inner product 
$\langle \cdot, \cdot \rangle$
on $\Bop(\Hch)$ by $\langle N , M \rangle= 2^{-1} {\rm Tr} N^{\dagger} M$
(half the Hilbert-Schmidt inner product),
then $\{ \Ebe_x \}_{x\in\cX}$ is an orthonormal basis with respect to
this inner product, and hence $P=P_{\cA}$ in Theorem~\ref{th:main}
is rewritten as
\[
 P(y) = \sum_{x\in \cX} | \langle \Ebe_{y}, A_x \rangle |^2.
\]
In fact, one sees that $P(y)$ has a physical meaning as follows.
If we define an inner product between 
$\mbm{n}=(n_{00},n_{01},n_{10},n_{11})$ and
$\mbm{m}=(m_{00},m_{01},m_{10},m_{11})$ by
$\langle \mbm{n},\mbm{m} \rangle  = 2^{-1}\sum_{z\in\cX} n_{z} m_{z}^*$,
then $\langle \xi(\mbm{n}),\xi(\mbm{m}) \rangle  
= \langle \mbm{n},\mbm{m} \rangle$,
so that we have
\[
 P(y) = \sum_{x\in \cX} | \langle \mbm{n}_{y}, \mbm{a}_x \rangle |^2,
\]
where $\xi(\mbm{n}_y)=\Ebe_y$.
Now, imagine we perform the orthogonal measurement 
$\{ 2^{-1} \mbm{n}_y^{\dagger} \mbm{n}_y \}_{y\in\cX}$ on
the system in the state (\ref{eq:chois_matrix2}).
Then, we obtain the result $y$ with probability
\begin{eqnarray*}
\lefteqn{\frac{1}{2} \mbm{n}_y \frac{1}{2} \sum_{x\in\cX}
\mbm{a}_x^{\dagger} \mbm{a}_x \mbm{n}_y^{\dagger} }\\
&=&\frac{1}{4} \sum_{x\in\cX} \mbm{n}_y \mbm{a}_{x}^{\dagger}
\mbm{a}_x \mbm{n}_y^{\dagger}\\
&=& \sum_{x\in\cX} |\langle \mbm{n}_y, \mbm{a}_x\rangle|^2\\
&=& P(y).
\end{eqnarray*}

Then, from the property of von Neumann entropy~\cite{wehrl},
$H(P)$ is not smaller than the von Neumann entropy of the state
(\ref{eq:chois_matrix2}) and equals it
when $\mbm{n}_x$ is proportional to $\mbm{a}_x$ for each $x\in\cX$, 
which is fulfilled
by setting $\ket{0}=\ket{b_0}$ and $\ket{1}=\ket{b_1}$ (and
$\omega=\tilde{\omega}$ for $\dmn>2$). 
To complete the proof, we have only to notice that
any unitary map 
preserves the entropy of the state
that it acts on, which implies $H(P)$ does not decrease
by preprocessing of applying $\Id\tnsr \cU$
to $[\Id\tnsr \cA](\ket{\Phi^+}\bra{\Phi^+})$.
\myQED

\subsection{Comparison of Bounds}

In this appendix, we prove (\ref{eq:bound_cmp}),
which states that our bound $1-H(P_{\cU\cA})$ is not smaller than
the previously known one $1-H_1(p')$,
and then, calculate $1-H_1(p')$ for the amplitude-damping channel
as an example.
Putting $\ket{b_0}=\ket{0}$ and $\ket{b_1}=\ket{1}$
(and hence viewing state vectors in terms of
the basis $\{ \ket{00},\ket{01},\ket{10},\ket{11}\}$),
we shall use the argument in the previous appendix,
which applies to general CP maps $\cA$ except the last paragraph.

First, we prove (\ref{eq:bound_cmp}).
As argued by Bennett {\em et al.}~\cite[p.~3830]{bennett96m},
every maximally entangled state can be represented, 
up to an overall phase factor, as
the transpose of $(u+\imu v, w+\imu z, -w+\imu z,u-\imu v)$
with $u,v,w,z$ real, i.e., as $(x,y,-y^*,x^*)^{\rm t}$,
where $xx^*+yy^*=1/2$.
Suppose $\ket{\eta}=x\ket{00}+y\ket{01}-y^*\ket{10}+x^*\ket{11}$
achieves the maximum in (\ref{eq:pprime}).
Then, putting $\mbm{\varvn}=\sqrt{2}(x^*,y^*,-y,x)$,
this maximum can be written as
\begin{eqnarray*}
\lefteqn{\frac{1}{2} \mbm{\varvn} \frac{1}{2} \sum_{\varx\in\cX}
\mbm{a}_\varx^{\dagger} \mbm{a}_\varx \mbm{\varvn}^{\dagger} }\\
&=& \sum_{\varx\in\cX} |\langle \mbm{\varvn}, \mbm{a}_\varx\rangle|^2\\
&=& \sum_{\varx\in\cX} |\langle U, A_\varx\rangle|^2 \\
&=& \sum_{\varx\in\cX} |2^{-1}{\rm Tr}  U^{\dagger} A_\varx |^2\\
&=& \sum_{\varx\in\cX} |\langle I, U^{\dagger} A_\varx \rangle|^2\\
&=& P_{\cU \cA}\big((0,0)\big)
\end{eqnarray*}
where $U=\xi(\mbm{\varvn})$, and $\cU(\rho)=U^{\dagger} \rho U$
(note that $U$ is unitary).
Hence, $1-H(P_{\cU\cA}) \ge 1-H_1\left(1-P_{\cU\cA}\big((0,0)\big)\right)$,
and the inequality is strict unless 
$P_{\cU\cA}\big((1,0)\big)=P_{\cU\cA}\big((0,1)\big)=P_{\cU\cA}\big((1,1)\big)$
by the property of the Shannon entropy $H$.
\myQED

{\em Example~4.}\/
We have calculated $1-H(P_{\cA})$ for the amplitude-damping channel
in Example~1.
For comparison, we compute $1-H_1(p')$ with (\ref{eq:pprime}) for this channel.
For the operator-sum representation in Example~1, we have 
$\mbm{a}_{(0,0)}=(1,0,0,\sqrt{1-\gamma})$ and
$\mbm{a}_{(1,0)}=(0,0,\sqrt{\gamma},0)$.
Hence, the maximized quantity in (\ref{eq:pprime})
can be calculated as
\[
\frac{1}{2} \mbm{\varvn} \frac{1}{2} \sum_{\varx=(0,0),(1,0)}
\mbm{a}_\varx^{\dagger} \mbm{a}_\varx \mbm{\varvn}^{\dagger} =
\gamma/4 + (1-\gamma+\sqrt{1-\gamma}) u^2 + (1-\gamma-\sqrt{1-\gamma}) v^2, 
\]
where $u={\rm Re}\, x$ and $v={\rm Im}\, x$.
From the normalization constraint $0\le u^2+v^2 \le 1/2$, it follows that
the maximum is $(2-\gamma+2\sqrt{1-\gamma})/4$ 
and hence, $p'=1-(2-\gamma+2\sqrt{1-\gamma})/4$.
\enexam

\subsection{Proof of Lemma~\protect\ref{lem:Preskill_bound}}

We employ the recovery operator $\cR \sim \{ {\cal O} \} \cup \{R_r\}$
constructed in the proof of Theorem~III.2 of \cite{KnillLaflamme97}
as well as the notation therein, where in the present case
their $\{ A_a \}$ are to be read $\{ \Ebe_x \}$.
Since the conditions (19) and (20) in Theorem~III.2 of \cite{KnillLaflamme97}
can be restated without referring to the code basis 
$\{ \ket{0_L },\dots, \ket{(K-1)_L} \}$
(see, e.g., \cite{knill96a,gottesman00a}),
we can assume $\ket{\psi}=\ket{0_L}$ without loss of generality.
Suppressing the superscript 
of $A_x^{(n)}$ and
using the relations $R_r = V_r \sum_{i} \ket{\nu_r^i} \bra{\nu_r^i}$
and $V_r\ket{\nu_r^i}=\ket{i_L}$~\cite{KnillLaflamme97}, we have
\begin{eqnarray*}
F(\psi) & = & 
  \sum_{r} \sum_{x} \bra{0_L} R_r A_x \ket{0_L} \bra{0_L} A_x^{\dagger} R_r^{\dagger} \ket{0_L}\\
  &=& \sum_{r}\sum_{x} \bra{\nu_r^0} A_x \ket{0_L}\bra{0_L} A_x^{\dagger} \ket{\nu_r^0}\\
  &=& \sum_{x} \bra{0_L} A_x^{\dagger} \Pi_0 A_x \ket{0_L},
\end{eqnarray*}
where we have put $\Pi_i=\sum_{r} \ket{\nu_r^i}\bra{\nu_r^i}$,
$0\le i \le K-1$.
Also we put $\Pi_{K}={\cal O}=I-\sum_{0\le i \le K-1}\Pi_i$.
Thus,
\begin{eqnarray*}
1-F(\psi) 
 & = & \sum_{1\le i\le K} \sum_{x} \bra{0_L} A_x^{\dagger} \Pi_i A_x \ket{0_L}\\
 & = & \sum_{1\le i \le K} \sum_{x}\sum_{y,z} 
       a_{xy}^*a_{xz} \bra{0_L} \Ebe_{y}^{\dagger} \Pi_i \Ebe_z \ket{0_L}\\
 & = & \sum_{1\le i \le K} \sum_{x}\sum_{y,z\in \Icr\cmple} 
       a_{xy}^*a_{xz} \bra{0_L} \Ebe_{y}^{\dagger} \Pi_i \Ebe_z \ket{0_L}\\
 & = & \sum_{1\le i \le K} \sum_{x} \bra{0_L} B_x^{\dagger} \Pi_i B_x \ket{0_L}\\
 & \le & \sum_{x} \bra{0_L} B_x^{\dagger} B_x \ket{0_L},
\end{eqnarray*}
where $B_x=\sum_{y\in \Icr\cmple} a_{xy}\Ebe_y$. \myQED

\subsection{Proof of Lemma~\protect\ref{lem:C_uniform}}

That $\crd{\Acn(0)}=0$ is trivial.
The lemma follows if we show 
that $\crd{\Acn(x)}=\crd{\Acn(y)}$ 
for any two distinct nonzero vectors $x$ and $y$.
This is because if it is so, putting $M=\crd{\Acn(x)}$, $x\ne 0$,
and counting the pair $(x,\Cso)$ such that $x\in\Cso^{\perp}$,
$\Cso\in\Acn$ and $x\ne 0$ in two ways, we will have
$(\dmn^{2n}-1)M=\crd{\Aso}(\dmn^{n+\kpr}-1)$.
To prove $\crd{\Acn(x)}=\crd{\Acn(y)}$, we use the Witt lemma,
which states that for a space $V$ with a nondegenerate (nonsingular)
symplectic form 
and subspaces $U$ and $W$ of $V$, if an isometry
(an invertible linear map that preserves the inner-product) 
$\alpha$ from $U$ to $W$
exists, then $\alpha$ can be extended to an isometry
from $\myF^{2n}$ onto itself~\cite[p.~81]{aschbacher},  
\cite[Theorem~3.9]{artin}.
First, note that any linear map 
from the space $\spn \{ x\}$ to $\spn \{ y\}$
preserve the symplectic inner product 
(\ref{eq:symplectic_form}), which always equals $0$
on these spaces. Among such maps, 
we choose the isometry $\alpha$ with $y=\alpha(x)$.
Then, by the Witt lemma,
$\alpha$ can be extended to $\myF^{2n}$.
Since
$\Cso \in \Aso(x)$ implies $\alpha(\Cso) \in \Aso(y)$,
we have $\crd{\Aso(x)} \ge \crd{\Aso(y)}$;
since
$\Cso \in \Aso(y)$ implies $\alpha^{-1}(\Cso) \in \Aso(x)$,
we have $\crd{\Aso(x)} \le \crd{\Aso(y)}$.
Hence, $\crd{\Aso(x)}=\crd{\Aso(y)}$, establishing the lemma.
\myQED

\end{document}